\documentclass[trackchanges]{aastex631}

\usepackage{CJK}
\usepackage{amsmath}
\usepackage{bm}
\shorttitle{3D magnetic switchbacks}
\shortauthors{Shi et al.}
\graphicspath{{./}{figures_new/}}

\begin{document}
\begin{CJK*}{UTF8}{gbsn}
\title{Analytic model and magnetohydrodynamic simulations of three-dimensional magnetic switchbacks}

\correspondingauthor{Chen Shi}
\email{cshi1993@ucla.edu}

\author[0000-0002-2582-7085]{Chen Shi (时辰)}
\affiliation{Department of Earth, Planetary, and Space Sciences, University of California, Los Angeles \\
Los Angeles, CA 90095, USA}
    
\author[0000-0002-2381-3106]{Marco Velli}
\affiliation{Department of Earth, Planetary, and Space Sciences, University of California, Los Angeles \\
Los Angeles, CA 90095, USA}

\author[0000-0001-8459-2100]{Gabor Toth}
\affiliation{Department of Climate and Space Sciences and Engineering, University of Michigan, Ann Arbor, MI 48109, USA}

\author[0000-0003-0393-9833]{Kun Zhang (张琨)}
\affiliation{Department of Earth, Planetary, and Space Sciences, University of California, Los Angeles \\
Los Angeles, CA 90095, USA}

\author[0000-0003-2880-6084]{Anna Tenerani}
\affiliation{Department of Physics, The University of Texas at Austin, \\
     TX 78712, USA}

\author[0000-0001-9570-5975]{Zesen Huang (黄泽森)}
\affiliation{Department of Earth, Planetary, and Space Sciences, University of California, Los Angeles \\
Los Angeles, CA 90095, USA}

\author[0000-0002-1128-9685]{Nikos Sioulas}
\affiliation{Department of Earth, Planetary, and Space Sciences, University of California, Los Angeles \\
Los Angeles, CA 90095, USA}

\author[0000-0001-5260-3944]{Bart van der Holst}
\affiliation{Department of Climate and Space Sciences and Engineering, University of Michigan, Ann Arbor, MI 48109, USA}



\begin{abstract}
Parker Solar Probe observations reveal that the near-Sun space is almost filled with magnetic switchbacks (``switchbacks'' hereinafter), which may be a major contributor to the heating and acceleration of solar wind. Here, for the first time, we develop an analytic model of an axisymmetric switchback with uniform magnetic field strength.
In this model, three parameters control the geometry of the switchback: height (length along the background magnetic field), width (thickness along radial direction perpendicular to the background field), and the radial distance from the center of switchback to the central axis, which is a proxy of the size of the switchback along the third dimension. 
We carry out three-dimensional magnetohydrodynamic simulations to investigate the dynamic evolution of the switchback.
Comparing simulations conducted with compressible and incompressible codes, we verify that compressibility, i.e. parametric decay instability, is necessary for destabilizing the switchback. 
Our simulations also reveal that the geometry of the switchback significantly affects how fast the switchback destabilizes. 
The most stable switchbacks are 2D-like (planar) structures with large aspect ratios (length to width), consistent with the observations.
We show that when plasma beta ($\beta$) is smaller than one, the switchback is more stable as $\beta$ increases. However, when $\beta$ is greater than one, the switchback becomes very unstable as the pattern of the growing compressive fluctuations changes.
Our results may explain some of the observational features of switchbacks, including the large aspect ratios and nearly constant occurrence rates in the inner heliosphere.
\end{abstract}



\section{Introduction} \label{sec:intro}
One of the most striking findings made by Parker Solar Probe (PSP) is that the nascent solar wind is almost filled with magnetic switchbacks (``switchbacks'' hereinafter) \citep{kasper2019alfvenic,bale2019highly}. Switchbacks are local polarity reversals of the radial magnetic field and are typically Alfv\'enic structures, with highly correlated velocity and magnetic field fluctuations and nearly constant magnetic field strength \citep{mcmanus2020cross,woolley2020proton,larosa2021switchbacks}. For a comprehensive description of switchbacks and their properties, please refer to the review paper \citep{raouafi2023parker} and references therein.

PSP observations have revealed that switchbacks significantly modify the properties of solar wind turbulence, highlighting their influential role in the solar wind dynamics \citep{de2020switchbacks,bourouaine2020turbulence,martinovic2021multiscale,shi2022patches}. 
Besides, as large-amplitude fluctuations, switchbacks are believed to be an important energy source for the heating and acceleration of the solar wind \citep{halekas2023quantifying}. Indeed, observations show that the plasma properties, including proton temperature and temperature anisotropy \citep{farrell2020magnetic,larosa2021switchbacks,woolley2021plasma,woodham2021enhanced,luo2023statistical,huang2023temperature,laker2024coherent}, are different inside and outside the switchbacks, with a possible enhancement of proton temperature inside the switchbacks. 


Although there is no doubt that understanding switchbacks is crucial for a complete understanding of the solar wind, how are the switchbacks generated and how do they propagate in the solar wind is still under debate. 
There are three widely accepted theories for the generation mechanisms of switchbacks, namely the interchange reconnection happening in the solar corona \citep{yamauchi2002relation,fisk2020global,zank2020origin,drake2021switchbacks,he2021solar,telloni2022observation,upendran2022formation,hou2023nature}, the velocity shear \citep{landi2006heliospheric,toth2023theory} that may arise due to coronal jets \citep{sterling2020coronal,magyar2021could,raouafi2023magnetic} or motion of the magnetic footpoint between source regions of fast and slow streams \citep{schwadron2021switchbacks}, and the natural evolution of Alfv\'en waves in the expanding solar wind \citep{belcher1971alfvenic,hollweg1974transverse,squire2020situ,shoda2021turbulent,mallet2021evolution}. 
However, which mechanism is dominant is still under debate. As different generation mechanisms take place at different radial distances to the Sun, it is thus important to study the radial evolution of the occurrence rate of switchbacks that may suggest which generation mechanism is most effective.
Unfortunately, no solid conclusion has been made on this topic either. 
\citet{mozer2020switchbacks} suggest that the switchback occurrence rate does not change much with radial distance, while other studies indicate a dependence of the occurrence rate on the duration of the switchback, i.e. the occurrence rate of larger switchbacks increases with the radial distance and the occurrence rate of shorter-duration switchbacks shows a decreasing trend \citep{tenerani2021evolution, jagarlamudi2023occurrence}.
Moreover, there is evidence showing that the occurrence rate of switchbacks is smaller inside the Alfv\'en radius than outside \citep{pecora2022magnetic,bandyopadhyay2022sub}. 
Hence, it is likely that a subset of the switchbacks observed in-situ are generated in the lower corona while the others are generated locally \citep{tenerani2021evolution}.
Another important question is how does a switchback evolve in the solar wind, because the dynamic evolution of switchbacks inevitably affects the characteristics of switchbacks observed in-situ, including their occurrence rates as well as their topology \citep{laker2021statistical}. Magnetohydrodynamic (MHD) simulations \citep{tenerani2020magnetic} show that a 2D Alfv\'enic switchback maintains stable for quite a long time (tens of Alfv\'en crossing time) and is eventually destabilized by the growth of density fluctuations, i.e. parametric decay instability (PDI). In addition, the Parker spiral \citep{johnston2022properties,squire2022properties} and Hall effect \citep{tenerani2023dispersive} also modify the evolution and dissipation of the switchbacks. 

The objective of this study is to numerically investigate the dynamic evolution of 3D switchbacks. The 3D effect has not been fully investigated in previous studies because of the difficulty to establish a 3D analytic model that can satisfy both the divergence-free condition and a uniform magnetic field strength.
\citet{squire2022construction} developed a time-integration method that grows the amplitude of the initially weak spherically-polarized magnetic field fluctuations (with normalized magnetic field fluctuation $\delta \bm{B}/ |\bm{B}| \approx 0.2$) while keeping the uniform-magnitude constraint and divergence-free condition. 
Here, we construct an analytic solution of a single 3D switchback, so that we are able control the geometry of the switchback and investigate the effect of the switchback's geometry on its evolution. 
In addition, we analyze how the plasma beta affects the destabilization process of the switchbacks.

The paper is organized as follows: In Section \ref{sec:analytic_form}, we describe in detail the analytic model of the 3D switchback. In Section \ref{sec:simulation}, we show the results of a series of 3D MHD simulations conducted based on the analytic model. In Section \ref{sec:summary}, we summarize the study and discuss possible future investigation.

\begin{figure}[htb!]
    \centering
    \includegraphics[width=\textwidth]{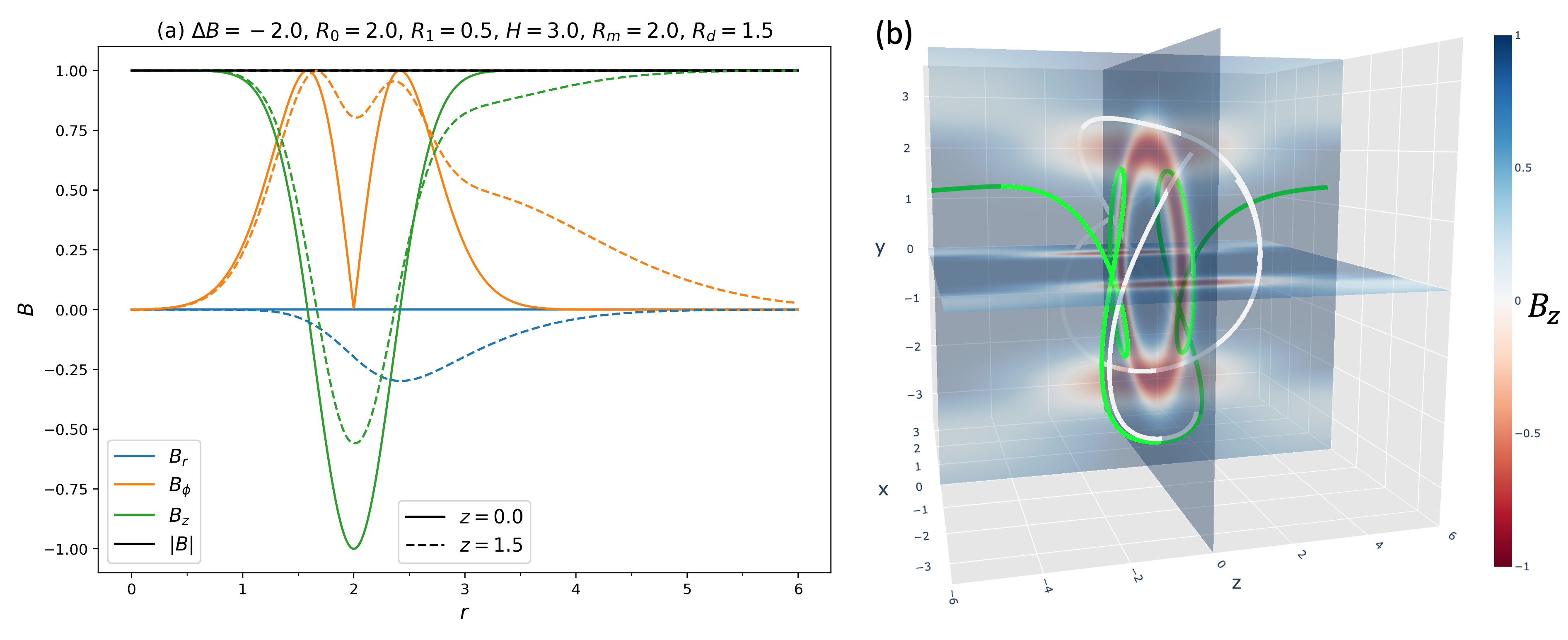}
    \caption{Visualization of the analytic model of the magnetic field described in Section \ref{sec:analytic_form}. The  model parameters are: $\Delta B/B_0 = -2$, $R_0 = 2$, $R_1=0.5$, $H=3$, $R_m = 2$, and $R_d = 1.5$. (a) Radial profiles of the $r$ (blue), $\phi$ (orange), $z$ (green) components, and magnitude (black) of the magnetic field. (b) Two magnetic field lines (green and white) over-plotted on three orthogonal surfaces which are color-coded with $B_z$. The background field is along $z$ direction.}
    \label{fig:1D_cut_fieldlines}
\end{figure}

\section{Analytic model of a 3D axisymmetric magnetic switchback}\label{sec:analytic_form}
We consider a cylindrical coordinate system $(r,\phi,z)$ and assume axisymmetric ($\partial/\partial \phi \equiv 0 $).
In this case, the magnetic field can be expressed by a scalar flux function $\psi(r,z)$ plus a $\phi$-component:
\begin{equation}
    \bm{B} = \frac{1}{r} \nabla \psi \times \hat{e}_{\phi} + B_\phi \hat{e}_{\phi} = - \frac{1}{r} \frac{\partial \psi}{\partial z} \hat{e}_{r} + \frac{1}{r}  \frac{\partial \psi}{\partial r} \hat{e}_{z} + B_\phi \hat{e}_{\phi},
\end{equation} 
which automatically satisfies the $\nabla \cdot \bm{B} =0 $ condition. 
We further decompose $\psi$ into a uniform background field $\psi_0 = \frac{1}{2} B_0  r^2$, which corresponds to $\bm{B_0} = B_0 \hat{e}_{z}$, and a fluctuation part $\psi_1(r,z)$. In order to have a switchback, we assume that $\psi_1$ has the form
\begin{equation}
    \psi_1(r,z) = \Delta B \exp \left(-\frac{z^2}{H^2} \right) f(r)
\end{equation}
where $\Delta B$ is the strength of the switchback, $H$ is the scale length of the switchback along $z$ direction, i.e. its ``height''. 
To confine the switchback within a region in radial direction, we further require
\begin{equation}\label{eq:equation_for_f}
    \frac{1}{r} \frac{d f}{dr} = \exp \left[ - \frac{(r-R_0)^2}{R_1^2} \right] 
\end{equation}
where $R_0$ is the radial location of the switchback's center, and $R_1$ is the radial size of the switchback, i.e. its ``width''. 
The above model of magnetic field is 2.5D instead of fully 3D as we are assuming symmetry in $\phi$. Nonetheless, the parameter $R_0$ serves as a proxy of the size of the switchback along the third dimension: The larger $R_0$ is, the more 2D-like, i.e. planar, the structure is.
In Appendix \ref{sec:app_model_Toth}, we present an alternative model which consists of trigonometric functions instead of Gaussian functions.

The solution of $f(r)$ to Equation (\ref{eq:equation_for_f}) is 
\begin{equation}
    f(r) = -\frac{1}{2}R_1^2 \left[ e^{- \frac{(r-R_0)^2}{R_1^2}} +\sqrt{\pi} \frac{R_0}{R_1} \mathrm{erf}\left(\frac{R_0-r}{R_1}\right) + C \right] 
\end{equation}
where $\mathrm{erf}(x)$ is the error function and the integral constant $C$ is chosen such that $f(0)=0$, i.e.
\begin{equation}
    C = - \left[ e^{-\frac{R_0^2}{R_1^2}}  + \sqrt{\pi} \frac{R_0}{R_1} \mathrm{erf}\left(\frac{R_0}{R_1}\right)\right]
\end{equation}
Thus, the analytic form of $B_r$ and $B_z$ are 
\begin{subequations}\label{eq:BrBz_analytic}
\begin{equation}\label{eq:Br_analytic}
     B_r(r,z) = - \Delta B \frac{zR_1^2}{r H^2} e^{-\frac{z^2}{H^2}} \left[ e^{- \frac{(r-R_0)^2}{R_1^2}} +\sqrt{\pi} \frac{R_0}{R_1} \mathrm{erf}\left(\frac{R_0-r}{R_1}\right) + C\right]
\end{equation}
\begin{equation}\label{eq:Bz_analytic}
   B_z(r,z) = B_0 + \Delta B \exp \left[ - \frac{(r-R_0)^2}{R_1^2} - \frac{z^2}{H^2}\right]
\end{equation}
\end{subequations}

Equation (\ref{eq:Bz_analytic}) indicates that the switchback is centered at $(r=R_0,z=0)$ and has a size $(R_1,H)$ in the $r-z$ plane. 
Ideally, $B_\phi$ can be used to maintain the constant-$|B|$ condition. 
However, Equation (\ref{eq:BrBz_analytic}) does not ensure $B_r^2 + B_z^2 \leq B_0^2$, and there are certain regions where the constant-$|B|$ condition cannot be preserved. 
The reason is that in the above model, $B_r \rightarrow 1/r$ for large $r$, which is not fast enough to maintain $B_r^2 + B_z^2 \leq B_0^2$. 
This also leads to a difficulty in setting up the boundary condition because we cannot directly implement periodic boundary conditions if $B_r$ does not decay to zero at the boundaries.
To overcome this difficulty, we modify the flux function with a predefined function $p(r)$ such that $B_r$ decays exponentially outside the switchback region and $B_z$ does not change in the z=0 plane:
\begin{equation}\label{eq:modified_fluxfunction}
    \tilde{\psi}(r,z) = p(r) \psi(r,z) + 
\left[1 - p(r) \right] \psi(r,0).
\end{equation}
For $p(r)$, we choose the following function
\begin{equation}\label{eq:pr_gaussian}
    p(r) = \left\{ \begin{array}{cc}
        \exp \left[ - \left(\frac{r - R_m}{R_d}\right)^2 \right] & r \geq R_m \\
        1 &  r<R_m
    \end{array} \right. 
\end{equation}
where $R_m$ is the radial location at which the modulation of the magnetic field starts, and $R_d$ is the scale length of the modulation. 
With the modified flux function, one can calculate the modified magnetic field
which results in the modified $B_r$ and $B_z$:
\begin{subequations}\label{eq:modified_Br_Bz_simple}
    \begin{equation}\label{eq:Br_modified}
        \tilde{B}_r = p(r) B_r(r,z)
    \end{equation}
    \begin{equation}
        \tilde{B}_z = p(r) B_z(r,z) + \left[1 - p(r) \right] B_z(r,0) + \frac{1}{r} \frac{dp}{dr} \left[ \psi(r,z) - \psi(r,0) \right]
    \end{equation}
\end{subequations}
that satisfy $\tilde{B}_r^2 + \tilde{B}_z^2 \leq B_0^2$ and hence $B_\phi = \left(B_0^2 - \tilde{B}_r^2 - \tilde{B}_z^2\right)^{1/2}$. 
We note that this modulation indeed does not change the magnetic field at the $z=0$ plane.
A different but equivalent approach to do the modulation is to assume $B_r$ is modified by a given function $p(r)$ (Equation (\ref{eq:Br_modified})) and $B_z$ is modified by an unknown function $q(r,z)$, and then solve $q(r,z)$ such that the modified magnetic field preserves the divergence-free condition.
This approach is described in Appendix \ref{sec:app_modulate_br}.
We neglect the tildes above $B_r$ and $B_z$ hereinafter for convenience. 

In Figure \ref{fig:1D_cut_fieldlines}, we visualize the above analytic model with parameters $\Delta B/B_0 = 2$, $R_0=2$, $R_1=0.5$, $H=3$, $R_m = 2$, and $R_d = 1.5$.
Panel (a) shows the radial profiles of the three components (colored) and magnitude (black) of the magnetic field. Solid lines are $z=0$, i.e. through the center of the structure, and dashed lines are $z=H/2$.
In panel (b), the three orthogonal surfaces are color-coded with $B_z$, and the green and white lines are two magnetic field lines. 
Here, the green line is connected to the background magnetic field, while the white line is confined within the switchback region. 
That is to say, in our model, not all of the field lines are connected to the Sun. In contrast, in-situ observations show that the magnetic field inside switchbacks is connected to the Sun \citep{balogh1999heliospheric,kasper2019alfvenic}. 
Thus, this model is a combination of switchbacks and flux ropes. However, we note that magnetic flux ropes are occasionally observed near the switchbacks by PSP \citep{choi2024series}. 
This phenomenon implies the possibility that reconnection takes effect either during the generation of the switchbacks \citep{drake2021switchbacks,he2021solar} or during the propagation of the switchbacks in the solar wind.
In Appendix \ref{sec:app_rex_switchbacks}, we briefly discuss how the reconnection between two switchbacks may generate a flux rope.

One limitation of the model is that, in the region $\left|z \right|>H$ and $R_m \leq r \lesssim R_m + R_d$, $B_z$ is smaller than $B_0$ due to the introduced modification (see panel (b) of Figure \ref{fig:1D_cut_fieldlines} and Figure \ref{fig:2D_cuts_Bz}). 
Consequently, there is a finite $B_\phi$ in this region, i.e. the background field is twisted azimuthally. This low-$B_z$ region extends infinitely in $z$. 
Actually, the $\nabla \cdot \bm{B} = 0$ condition means it is impossible to find a switchback model that has finite extents in both $r$ (the $x-y$ plane) and $z$. 
The proof is simple: Suppose that the fluctuation part of the field totally vanishes outside a finite radial distance $R_b$ in the $x-y$ plane and consider a cylinder with $r=R_b$ and $0 \leq z \leq H_b$. Clearly there is no magnetic flux across the side surface $r=R_b$, 
hence the divergence-free condition writes as 
\begin{equation}
    \iint_{z=H_b,r \leq R_b} B_z = \iint_{z=0,r \leq R_b} B_z.
\end{equation}
As the perturbation in $B_z$ at the $z=0$ surface is negative definite, we immediately see that $B_z(z=H_b) \equiv B_0$ cannot satisfy the above equation, i.e. for any horizontal plane $z=H_b$, there must be some regions where $B_z<B_0$.

\begin{table}[htb!]
\begin{center}
\begin{tabular}{c|c|c|c|c|c|c|c|c|c}
  \hline
   Run & $\Delta B / B_0$  & $R_0$  & $R_1$ & $H$ & $R_m$ & $R_d$ & $P_0$ & $L_x,L_y,L_z$ & $n_x,n_y,n_z$  \\
   \hline 
   0, 0IC  & -2 & 2 & 0.5 & 3 & 2 & 1.5 & 1.0  & 24,24,24 & 512,512,512  \\
   \hline
   1  & -2 & 4 & 0.5 & 3 & 4 & 1.5 & 1.0  & 48,48,24 & 1024,1024,512  \\
   \hline
   2  & -2 & 2 & 0.5 & 6 & 2 & 1.5 & 1.0  & 24,24,48 & 512,512,1024  \\
   \hline
   3  & -2 & 2 & 0.5 & 1.5 & 2 & 1.5 & 1.0  & 24,24,12 & 512,512,256  \\
   \hline
   4  & -2 & 2 & 0.5 & 3 & 2 & 1.5 & 0.75  & 24,24,24 & 512,512,512  \\
   \hline
   5  & -2 & 2 & 0.5 & 3 & 2 & 1.5 & 0.5  & 24,24,24 & 512,512,512  \\
   \hline
   6  & -2 & 2 & 0.5 & 3 & 2 & 1.5 & 0.2  & 24,24,24 & 512,512,512  \\
   \hline
   7  & -2 & 2 & 0.5 & 3 & 2 & 1.5 & 0.1  & 24,24,24 & 512,512,512  \\
   \hline
   8  & -2 & 2 & 0.5 & 3 & 2 & 1.5 & 0.05  & 24,24,24 & 512,512,512  \\
   \hline
\end{tabular}
\caption{Simulation parameters. Additional low-resolution runs to verify the grid-convergence are not listed here (see the text).}
\label{tab:parameters}
\end{center}
\end{table}

\section{Numerical simulations}\label{sec:simulation}
\subsection{Simulation setup}
Based on the analytic model described in Section \ref{sec:analytic_form}, we conduct a series of 3D MHD simulations using a pseudo-spectral MHD code \citep{shi2019fast,shi2020propagation,shi2022influence,tenerani2023dispersive,shi2023evolution}. 
The code evolves the MHD equation set in conservation form on a Cartesian grid.
The initial condition comprises a uniform density $\rho_0 \equiv 1$, a uniform thermal pressure $P_0$, magnetic field as described in Section \ref{sec:analytic_form} with $B_0 = 1$, and velocity that is anti-correlated with the magnetic field ($\bm{V} = -\bm{B}/\sqrt{ \rho_0}$). We note that the code uses normalized quantities such that permeability $\mu_0$ is eliminated. 

\begin{figure}[htb!]
    \centering
    \includegraphics[width=\hsize]{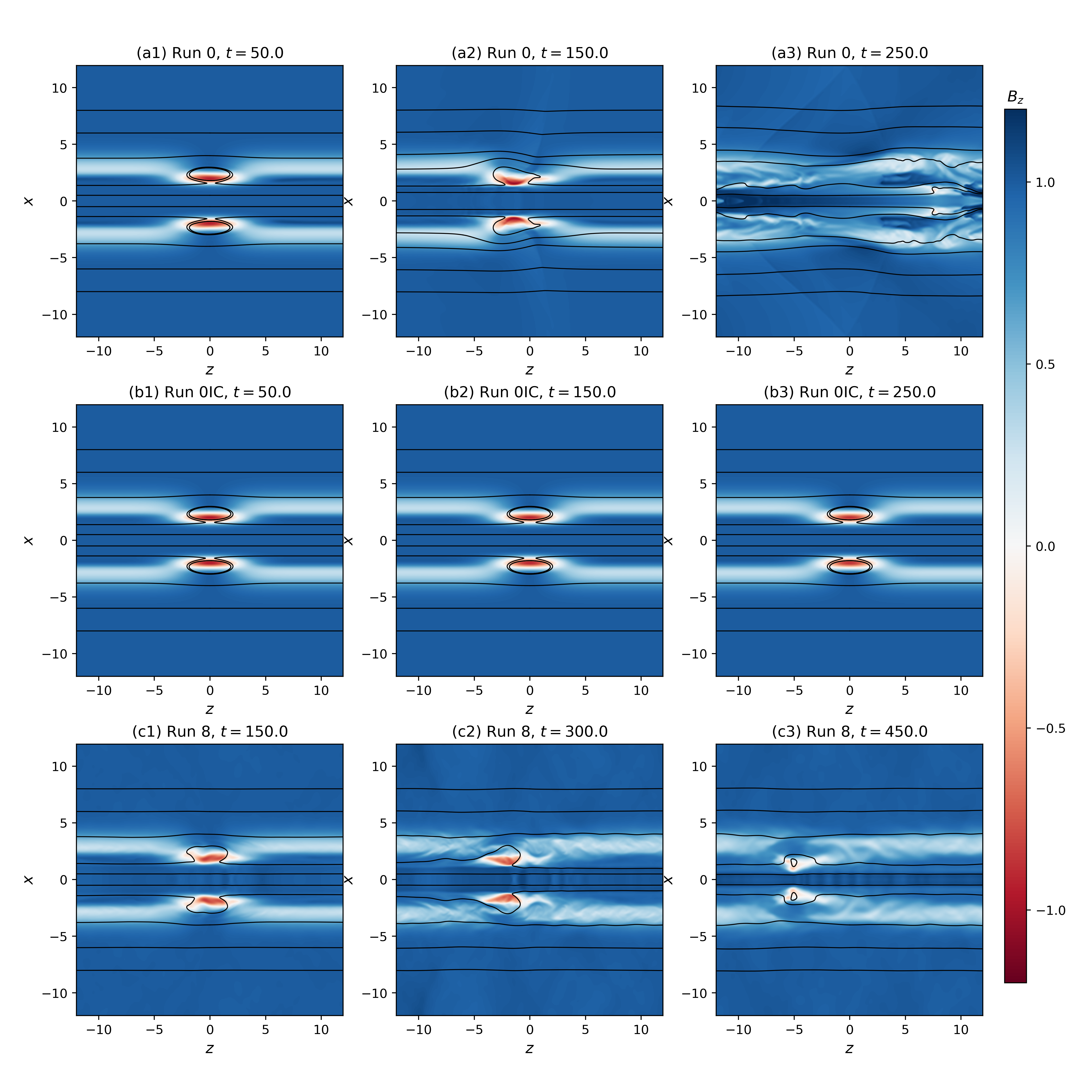}
    \caption{Snapshots of the $x-z$ plane across the central axis ($y=0$) color-coded with $B_z$ for Runs 0 (a,top), 0IC (b,middle), and 8 (c,bottom). Black curves are magnetic field lines projected on the 2D plane.}
    \label{fig:2D_cuts_Bz}
\end{figure}

The parameters for the conducted simulations are summarized in Table \ref{tab:parameters}. In all the runs listed in Table \ref{tab:parameters}, we use the same spatial resolution $\Delta = 0.047$ along all dimensions. 
In addition, several fixed parameters include the switchback strength $\Delta B/B_0 \equiv  2$, the width of the switchback $R_1 \equiv 0.5$, and the parameters for modification of the magnetic field outside the switchback region: $R_m \equiv R_0$, $R_d \equiv 1.5$. 
We do not implement explicit resistivity and viscosity. Instead, we apply a numerical filter in the wave-vector space, which is similar to the filters usually adopted in compact-finite-difference schemes \citep{lele1992compact}, to all quantities after each time step to maintain numerical stability. For a detailed description of the filter, please refer to equations (16,17) of \citet{shi2020propagation}. 
To verify grid convergence of the simulations, several low-resolution runs (not listed in Table \ref{tab:parameters}) based on Run 0 are conducted, including one run with $n_z=256$, one run with $n_x=n_y=256$, one run with $n_x=n_y=n_z=256$, and one run with $n_x=n_y=n_z=256$ and weaker numerical filter. We have verified that the linear growth rate of the instability is not affected by the resolution and the filter strength.

Compared with Run 0, Run 1 has doubled $R_0$, Run 2 has doubled $H$, and Run 3 has one half of $H$, so that we can investigate the effect of the geometry on the switchback evolution. 
We also carry out one run, labeled as Run 0IC, with exactly the same initial condition with Run 0, but using incompressible version of the simulation code.
For Runs 0-3, $P_0 = 1.0$ so that the plasma beta is $\beta = 2.0$. To investigate how $\beta$ influences the evolution, we conduct Runs 4-8 which have the same initial configuration with Run 0 but different values of $P_0$ so that $\beta$ varies in the range 0.1-2.0.

\subsection{Results}

\begin{figure}[htb!]
    \centering
    \includegraphics[width=\hsize]{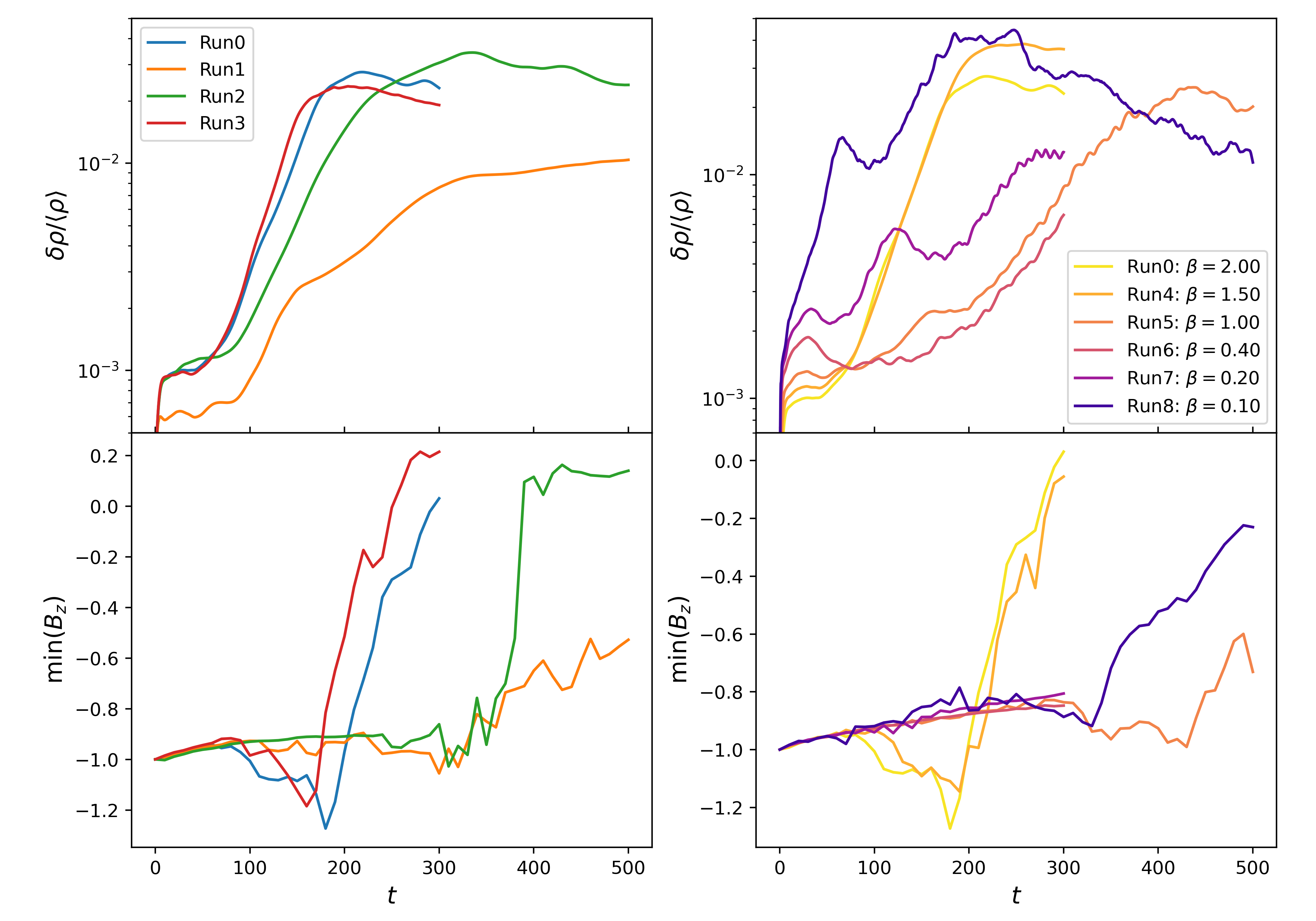}
    \caption{Time evolution of (top) the normalized density fluctuations $\delta \rho / \left< \rho \right>$ where $\delta \rho$ is the root-mean-square of the density and $\left<\rho \right>$ is the average density and (bottom) the minimum value of $B_z$ in the simulation domain. In the left column, we show the results for Runs 0, 1, 2, 3. In the right column  we show the results for Runs 0, 4, 5, 6, 7, 8.}
    \label{fig:rmsrho_minBz}
\end{figure}

In Figure \ref{fig:2D_cuts_Bz}, we show the snapshots of Runs 0 (top), 0IC (middle), \& 8 (bottom) color-coded with $B_z$. The plots are 2D slices in $x-z$ plane, across the central axis ($y=0$). 
The black lines are magnetic field lines projected on the 2D plane.
In Run 0IC, the switchback remains stable until the end of the simulation ($t=300$), while in Run 0 and Run 8, the switchback remains stable for a certain time but eventually destabilizes.
These results indicate that compressibility is a necessary condition for the destabilization of the switchbacks, verifying the conclusion of \citet{tenerani2020magnetic}, i.e. parametric decay instability is the major mechanism in the destabilization of the switchbacks. .

In Figure \ref{fig:rmsrho_minBz}, we plot the time evolution of the normalized density fluctuation ($\delta \rho / \left< \rho \right>$) (top) and minimum $B_z$ in the simulation domain (bottom) in different runs. 
Here $\delta \rho$ is the root-mean-square and $\left<\rho \right>$ is the average of density.
Let us first focus on the left column, which shows results of Runs 0 (blue), 1 (orange), 2 (green), and 3 (red). Apparently, the density fluctuations grow exponentially over certain time intervals in all the Runs before they saturate. 
As the density fluctuations grow, minimum $B_z$ gradually decreases below $-1$. After the density fluctuations saturate, minimum $B_z$ starts to rise rapidly and eventually becomes positive. 
That is to say, the magnetic field is first compressed as the PDI grows and then destabilizes rapidly after the PDI saturates.
We note that, due to the numerical diffusivity, min($B_z$) gradually increases with time if there is no compressibility or the growth rate of PDI is small.
Comparing Runs 0, 2, and 3, which have different values of $H$ (length/height of the switchback), we can see that the growth rate of the PDI is largest in Run 3 and smallest in Run 2, indicating that longer switchbacks are more stable than the shorter ones, consistent with the PSP observations which show that the switchbacks in the solar wind are typically very long structures \citep{horbury2020sharp,laker2021statistical}.
Comparing Run 0 and Run 1, we see that the switchback is much more stable in Run 1, in which $R_0$, the size of the structure along the third dimension, is larger. That is to say, the switchback is more stable if it is more 2D-like or planar.
In conclusion, Runs 0-3 show that the most stable switchbacks have 2D-like structures and large aspect ratios (length to width).

\begin{figure}[htb!]
    \centering
    \includegraphics[width=\hsize]{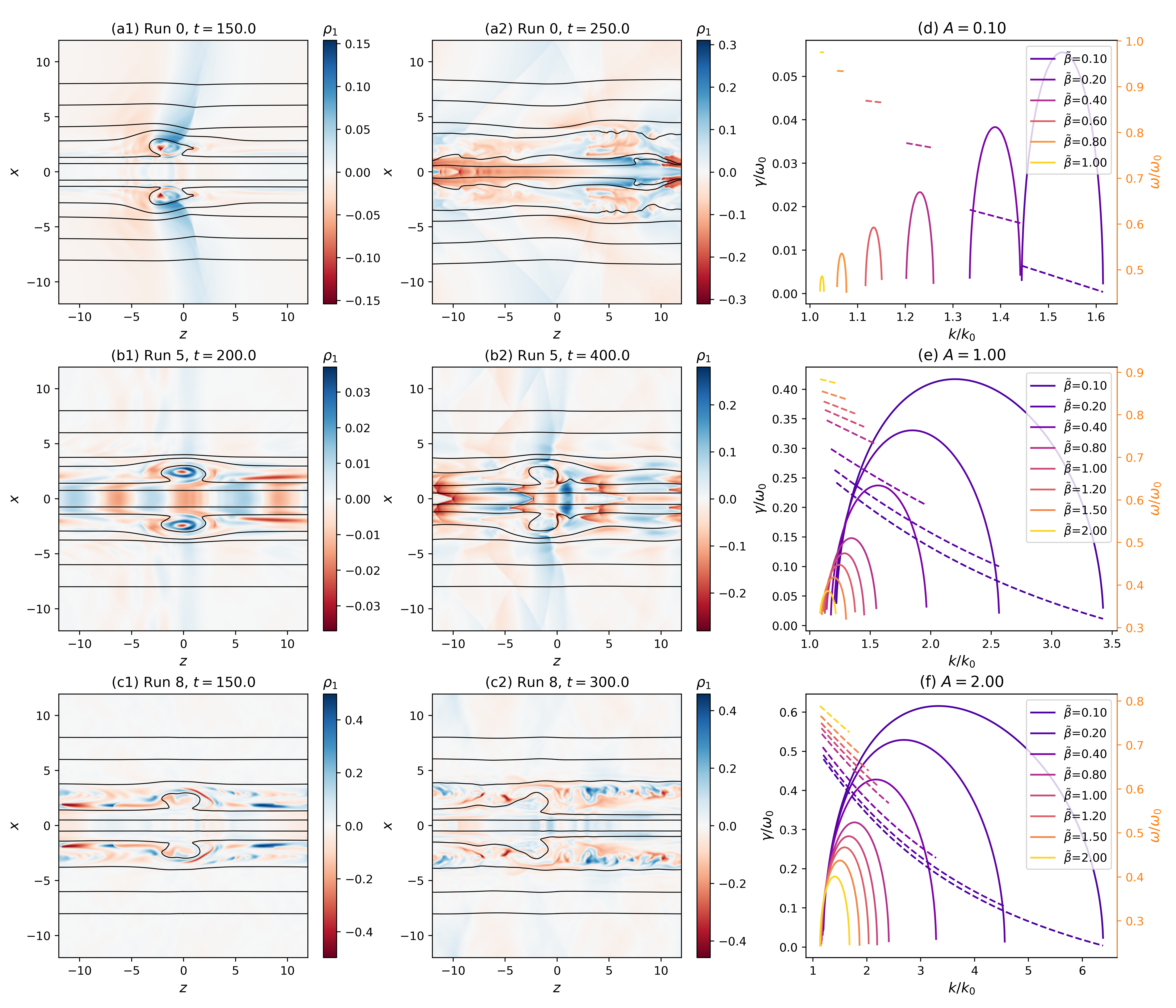}
    \caption{Panels (a1)-(c2): Snapshots of the $x-z$ plane across the central axis ($y=0$) color-coded with $\rho_1$ for Runs 0 (a,top), 5 (b,middle), and 8 (c,bottom) where $ \rho_1 = \rho - \left< \rho \right>$ is the density fluctuation. Black curves are magnetic field lines projected on the 2D plane. Panels (d)-(f): Linear growth rate (solid) and frequency (dashed) of the parametric decay instability for a monochromatic circularly-polarized Alfv\'en wave with amplitude (d) $A=0.1$, (e) $A=1.0$, and (f) $A=2.0$. In each panel, different curves correspond to different values of $\tilde{\beta} = C_s^2/V_A^2$, where $C_s$ and $V_A$ are the sound speed and Alfv\'en speed of the background field. $\omega_0$, $k_0$ are the frequency and wavenumber of the mother Alfv\'en wave. $A = \delta B / B_0$ is the amplitude of the mother wave normalized to the background magnetic field.}
    \label{fig:2D_cuts_rho1}
\end{figure}

Then we investigate the effect of plasma beta by inspecting the right column of Figure \ref{fig:rmsrho_minBz}. From the time profiles of $\delta \rho/ \left< \rho \right>$, we see that the evolution in runs with $\beta > 1$ is quite different from the evolution in runs with $\beta \leq 1$. 
The growth rates of the instability are almost identical in Run 0 ($\beta = 2.0$) and Run 4 ($\beta = 1.5$). 
When $\beta $ drops to unity (Run 5), the growth rate decreases significantly. As we further decrease $\beta$ to 0.4 (Run 6), the growth rate does not change much compared with Run 5. 
Eventually, as $\beta$ continues to drop (Run 8, $\beta = 0.1$), the growth rate starts to increase and becomes comparable to $\beta > 1$. 
Although the growth rates of the instability are comparable in Runs 0, 4, and 8, the disturbance of the magnetic field is much slower in Run 8 than in Runs 0 \& 4 (bottom right panel), implying that the characteristics of the density fluctuation may be very different depending on plasma beta.

It is well known that the growth rate of PDI depends on $\beta$ \citep{derby1978modulational,jayanti1993parametric}. For a monochromatic, circularly polarized  Alfv\'en wave, the dispersion relation of PDI is described by the equation \citep{derby1978modulational,tenerani2013parametric,reville2018parametric}
\begin{equation}\label{eq:pdi_linear}
  \left( \omega + k + 2 \right) \left( \omega + k - 2 \right) \left( \omega - k \right)  \left( \omega^2 - \tilde{\beta} k^2 \right) - A^2 k^2 \left( \omega^3 + k \omega^2 -3 \omega + k \right) = 0
\end{equation}
where $\omega$ and $k$ are frequency and wavenumber of the density fluctuation normalized to the frequency and wavenumber of the mother wave, $A = \delta B / B_0$ is the amplitude of the mother wave normalized to the background magnetic field, and $\tilde{\beta} = C_s^2/V_A^2$ where $C_s$ and $V_A$ are the sound speed and Alfv\'en speed of the background plasma.
Note that we use the symbol $\tilde{\beta}$ instead of $\beta$ since $\tilde{\beta} = 5/6 \times \beta$ assuming an adiabatic index of $5/3$.
In panels (d)-(f) of Figure \ref{fig:2D_cuts_rho1}, we plot the solved dispersion relation based on Equation (\ref{eq:pdi_linear}) for (d) $A=0.1$, (e) $A=1.0$, and (f) $A=2.0$ respectively. 
In each panel, the solid lines are growth rate and the dashed lines are frequency. Colors of the lines correspond to different values of $\tilde{\beta}$.
Clearly, regardless of the wave amplitude $A$, the growth rate decreases with $\tilde{\beta}$ and the wavenumber corresponding to the maximum growth rate decreases toward $k_0$. 
As pointed out by \citep{jayanti1993parametric}, for $\beta>1$, the instability is more like a beat instability as the wavenumber and frequency of the generated sound wave are close to those of the mother wave. 
In this case, the secondary transverse fluctuation is dominated by the forward propagating Alfv\'en wave with wavenumber close to $2k_0$.
In panels (a1)-(c2), we show the snapshots color-coded with $\rho_1$ in Runs 0 (a, $\beta =2.0$), 5 (b, $\beta =1.0$), and 8 (c, $\beta =0.1$) with the magnetic field lines. Here $\rho_1 = \rho - \left< \rho \right>$ is the density fluctuation.
We can see that, for $\beta=2.0$, during the linear stage (panel (a1)), the density fluctuation is concentrated within the switchback region. 
For small $\beta$($=0.1$) (panels (c1,c2)), the density fluctuation has shorter wavelength and is distributed along the $z$-direction, similar to that observed in 2D simulations conducted by \citet{tenerani2020magnetic} (see their Figure 6). 
For intermediate $\beta$ (Run 5, $\beta = 1.0$), we see growth of density fluctuations both inside and outside the switchback, and the wavelength of the growing mode is larger than Run 8. 
These results are qualitatively consistent with the linear theory. However, a major discrepancy is that, in our simulations, the growth rate of the instability for $\beta>1$ is larger than $\beta < 1$ except for very small $\beta$ ($\beta=0.1$), while in the linear theory, the maximum growth rate monotonically decreases with $\beta$.
We point out that, the simulated switchback is not a simple monochromatic, circularly-polarized, one-dimensional Alfv\'en wave, so the stability of this structure is likely to deviate from linear theory described by Equation (\ref{eq:pdi_linear}) \citep{malara1996parametric}.

\section{Summary}\label{sec:summary}
In this study, we develop an analytic model of an axisymmetric 3D magnetic switchback with uniform magnetic field strength. A series of 3D MHD simulations are conducted to investigate the evolution of the switchback. We focus on how the geometry and plasma beta modify the switchback evolution. The major conclusion of the simulations is
\begin{itemize}
    \item Compressibility is a necessary condition for the destabilization of the switchbacks \citep[see also][]{tenerani2020magnetic}. The switchback is stable in the incompressible-MHD simulation.
    \item The most stable switchbacks are  2D-like structures with large aspect ratios, i.e. planar structures elongated along the background magnetic field direction. 
    \item The plasma beta plays an important role in the destabilization of switchbacks. For $\beta \ll 1$, the structure is unstable due to the classic parametric decay instability. $\beta \lesssim 1$ is the most stable regime for the structure. For $\beta > 1$, the structure becomes unstable again due to the beat instability \citep{jayanti1993parametric}. Destabilization of the magnetic field in this case is much faster than $\beta < 1$.
\end{itemize}
Our results may explain why the switchbacks observed by PSP are mostly long structures \citep{horbury2020sharp,laker2021statistical}. 
Moreover, the $\beta$-dependence of the stability shown by the simulations suggests a scenario where the occurrence rate of switchbacks in the inner heliosphere ($\beta \lesssim 1$) does not change significantly with radial distance to the Sun \citep{mozer2020switchbacks,pecora2022magnetic}, but drops significantly when $\beta$ grows to greater than unity during the propagation of the solar wind.
This may explain why much less switchbacks are observed beyond 1 AU, e.g. Ulysses observation \citep[e.g.][]{balogh1999heliospheric}, than inside 1 AU.
An investigation of the relation between the occurrence rate of switchbacks and the plasma beta will be necessary to verify this point.
Last, we note that our simulations start with a switchback that is already generated. An important question is whether and how different generation mechanisms lead to different structures of the switchbacks.
However, this is beyond the scope of the current study. Future numerical and observational studies are necessary to answer this question.

\begin{acknowledgments}
This work is supported by NSF SHINE \#2229566 and NASA ECIP \#80NSSC23K1064. The numerical simulations are conducted on Extreme Science and Engineering Discovery Environment (XSEDE) EXPANSE (allocation No. TG-AST200031.) at San Diego Supercomputer Center (SDSC), which is supported by National Science Foundation grant number ACI-1548562 \citep{Townsetal2014}, and the Derecho: HPE Cray EX System (https://doi.org/10.5065/qx9a-pg09) of Computational and Information Systems Laboratory (CISL), National Center for Atmospheric Research (NCAR) (allocation No. UCLA0063). We thank Dr. Jaye L Verniero for very inspiring discussions.
\end{acknowledgments}

%



\software{Matplotlib \citep{Hunter2007Matplotlib}, NumPy \citep{harris2020array}
          }



\appendix
\section{An alternative analytic model of the switchback}\label{sec:app_model_Toth}
In Section \ref{sec:analytic_form}, we introduced an analytic model of the switchback (Equation (\ref{eq:BrBz_analytic})) based on Gaussian functions. 
Here, we show an alternative model which is based on simpler trigonometric functions and is more compact spatially. The definition of the parameters $\Delta B$, $R_0$, $R_1$, $H$, $R_m$, $R_d$ remains the same with the main text. 

The background field is still $\psi_0 = \frac{1}{2} B_0 r^2$. For the fluctuation part, we write
\begin{equation}
    \psi_1(r,z) = \Delta B \times g(z) f(r)
\end{equation}
where 
\begin{equation}
   g(z) =  \left\{ 
   \begin{array}{cc}
       \cos^2 \left( \frac{\pi}{2} \frac{z}{H} \right) & \left|z \right| \leq H  \\
       0 & \left| z \right| > H
   \end{array}  \right.
\end{equation}
We define $s=  R_1/\pi$, $x=(r-R_0)/2s$, and write $f(r)$ as
\begin{equation}
    f(r) = s^2 \left[x^2 + x \sin(2x) + \cos(2x)/2 \right] + s R_0 \left[ x + \sin(2x)/2 \right] + C, \quad   \left|r-R_0\right| \leq R_1
\end{equation}
The constant $C$ is such that $f(R_0 - R_1) = 0$. We require $f(r > R_0 + R_1) \equiv f(R_0 + R_1)$ and $f(r < R_0 - R_1) \equiv 0$.
One can easily show that the above $f(r)$ and $g(z)$ lead to 
\begin{equation}
    B_z(r,z) = B_0 + \Delta B \cos^2\left( \frac{\pi}{2} \frac{z}{H} \right) \cos^2 \left(\frac{\pi}{2} \cdot \frac{r-R_0}{R_1} \right), \quad \left| z \right| \leq H, \, \left|r-R_0\right| \leq R_1
\end{equation}
Similar to the model described in Section \ref{sec:analytic_form}, we need to modify the magnetic field so that $B_r$ vanishes outside a certain radius.
Here, we set the modification function $p(r)$ as
\begin{equation}
    p(r) = 
    \left\{
    \begin{array}{cc}
         1 & r< R_m  \\
         \cos^4 \left(\frac{\pi}{2} \cdot \frac{r-R_m}{R_d}\right) &  \quad R_m \leq r \leq R_m + R_d \\
         0 & r > R_m + R_d
    \end{array}
    \right.
\end{equation}
The finalized flux function is then described by Equation (\ref{eq:modified_fluxfunction}).
Based on experiments, we find that $R_m \leq R_0$ is necessary if $\Delta B = -2 B_0$, and $R_m < R_0 + R_1$ for $\Delta B > -2 B_0$.

\section{An alternative approach to modulate the magnetic field outside the switchback region}\label{sec:app_modulate_br}
In this section, we describe an approach that is equivalent to Equation (\ref{eq:modified_fluxfunction}) but directly modifies the magnetic field instead of the flux function. Suppose we have a predefined function $p(r)$ so that the modified $B_r$ is 
\begin{equation}
    \Tilde{B}_r (r,z) = p(r) B_r(r,z).
\end{equation}
Obviously, $B_z$ also needs to be modified by a function $q(r,z)$
\begin{equation}
    \Tilde{B}_z(r,z) = q(r,z) B_z(r,z)
\end{equation}
in order to preserve the divergence-free condition $\nabla \cdot \bm{\Tilde{B}} = 0$, which leads to an equation for $q(z)$ at a fixed $r$:
\begin{equation}\label{eq:equation_for_q_general}
    \frac{d q(z)}{d z} + P(z) q(z) = Q(z)
\end{equation}
where 
\begin{displaymath}
    P(z) = \frac{1}{B_z} \frac{\partial B_z}{\partial z}, \quad Q(z) = \frac{p}{B_z} \frac{\partial B_z}{\partial z} - \frac{B_r}{B_z} \frac{\partial p}{\partial r}.
\end{displaymath}
We note that $\nabla \cdot \bm{B} = 0$ is used to simplify the equation.
Equation (\ref{eq:equation_for_q_general}) is a typical first-order linear ordinary differential equation, and has a solution of the following form
\begin{equation}
    q(z) = \frac{1}{I(z)} \left[ C_q + \int_{0}^{z} I(z^\prime) Q(z^\prime) dz^\prime \right]
\end{equation}
where $C_q$ is an integral constant and
\begin{displaymath}
    I(z) = \exp \left[ \int_{0}^z P(z^\prime) dz^\prime \right] = \frac{B_z(z)}{B_z(0)}
\end{displaymath}
We further require $q(z=0) \equiv 1$ so that $B_z$ is not modified at the central plane $z=0$, which then leads to $C_q = 1$. 
Hence, the final form of $q(z)$ is
\begin{equation}\label{eq:qz_final}
    q(z) = \frac{B_z(0)}{B_z(z)} \left\{1 + \frac{1}{B_z(0)}\int_{0}^{z} \left[ p \frac{\partial B_z}{\partial z} - B_r \frac{\partial p}{\partial r}\right] dz^\prime \right\}
\end{equation}
Since we assume $p(r)$ is not a function of $z$, the integral on the right-hand-side of the above equation can be calculated easily:
\begin{equation}\label{eq:qz_integral_part1}
    \int_{0}^{z} p \frac{\partial B_z}{\partial z} dz^\prime = p [B_z(z) - B_z(0)] = p \times  \Delta B \exp \left[ - \frac{(r-R_0)^2}{R_1^2}\right] \left( e^{-\frac{z^2}{H^2}} - 1\right)
\end{equation}
and 
\begin{equation}\label{eq:qz_integral_part2}
\begin{aligned}
    \int_{0}^{z} B_r \frac{\partial p}{\partial r} dz^\prime & =  \frac{d p}{d r} \times \Delta B \frac{R_1^2}{r} \left[ e^{- \frac{(r-R_0)^2}{R_1^2}} +\sqrt{\pi} \frac{R_0}{R_1} \mathrm{erf}\left(\frac{R_0-r}{R_1}\right) + C\right] \times \frac{1}{2}\left(e^{-\frac{z^2}{H^2}} - 1 \right)
\end{aligned}
\end{equation}
One can verify that the result acquired using the above procedure is exactly the same with Equation (\ref{eq:modified_Br_Bz_simple}).

\section{A possible mechanism that generates flux ropes near switchbacks}\label{sec:app_rex_switchbacks}
\begin{figure}[htb!]
    \centering
    \includegraphics[width=0.75\hsize]{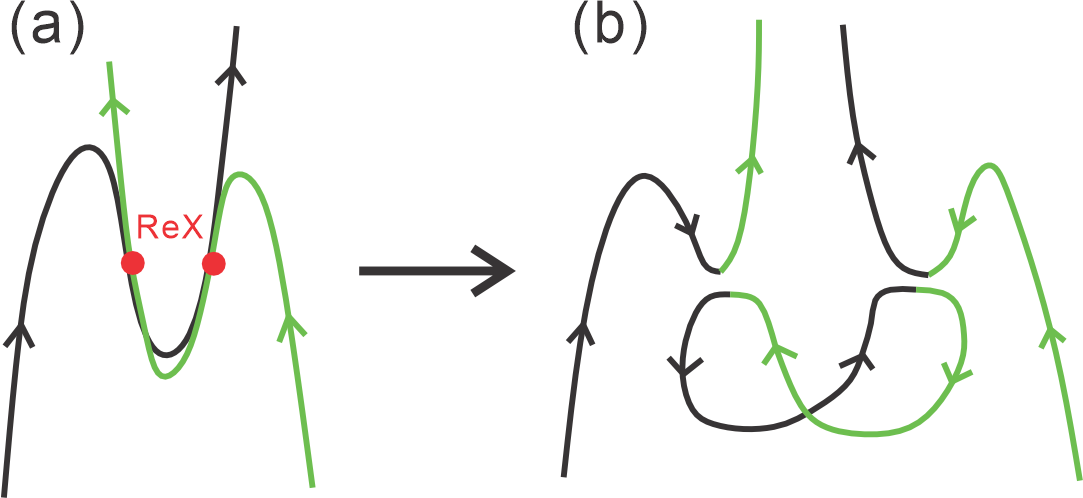}
    \caption{Illustration of two interacting switchbacks. (a) Reconnection happens at two different locations marked by the red dots. (b) Two new switchbacks and a flux rope are generated.}
    \label{fig:switchback_reconnection}
\end{figure}
In all the analytic models of switchbacks (current study and previous study by \citet{tenerani2020magnetic}), closed field lines, i.e. flux ropes, are inevitably included as there is no straightforward way to construct a model with open magnetic field lines only. 
The existence of flux ropes around switchbacks indicates prior reconnection events. However, we emphasize that switchbacks accompanied by flux ropes are not necessarily generated by interchange reconnection \citep{drake2021switchbacks,he2021solar}.
One possible mechanism is the interaction between two switchbacks during the propagation of the solar wind, as illustrated by Figure \ref{fig:switchback_reconnection}.
When the two switchbacks collide, reconnection happens at two different locations (a). At each reconnection site, the inverse part of one field line reconnects with the normal part of the other filed line.
After the reconnection (b), two new switchbacks and a flux rope are generated.
We note that, in practice, it may not be easy to trigger reconnection in switchbacks as they are highly Alfv\'enic. Nonetheless, reconnection is occasionally observed near the switchback boundaries \citep{froment2021direct}.
Thus, the mechanism shown by Figure \ref{fig:switchback_reconnection} should not be excluded.


\bibliography{references}{}
\bibliographystyle{aasjournal}



\end{CJK*}
\end{document}